\documentclass[superscriptaddress,twocolumn,showpacs,amsmath,amssymb]{revtex4}
\usepackage{graphicx}% Include figure files
\usepackage{dcolumn}% Align table columns on decimal point
\usepackage{bm}% bold math

\renewcommand{\vec}[1]{\mbox{\boldmath $#1$}}

\begin{document}

\preprint{}

\title{Deformation parameter for diffuse density}% Force line breaks with \\

\author{K. Hagino}
\affiliation{
Department of Physics, Tohoku University,
Sendai 980-8578, Japan}

\author{N. W. Lwin}
\affiliation{
Department of Physics, Tohoku University,
Sendai 980-8578, Japan}

\author{M. Yamagami}
\affiliation{
Radioactive Isotope Physics Laboratory, RIKEN, Wako, 
Saitama 351-0198, Japan}

\date{\today}% It is always \today, today,
             %  but any date may be explicitly specified

\begin{abstract}

In extracting deformation parameters 
from multipole moments for deformed nuclei, one commonly 
uses the formulas which are based on a sharp-cut density distribution. 
We discuss a possible ambiguity for this procedure and 
clarify the role of diffuseness parameter of the density
distribution. 
For this purpose, we use a deformed Woods-Saxon 
density as well as a density distribution 
obtained from the self-consistent 
relativistic mean-field (RMF) model. 
We show that the formula using a root-mean-square radius instead 
of a sharp-cut radius requires 
a large correction 
due to a finite surface diffuseness parameter even for stable nuclei. 
An implication to neutron-rich nuclei is also discussed. 

\end{abstract}

%\keywords{Suggested keywords}%Use showkeys class option if keyword
                              %display desired
\maketitle

%\section{Introduction}
%26.2.06

Deformation of a density distribution in the intrinsic frame 
is one of the most important concepts in nuclear physics. 
It is characterized by a finite value of intrinsic quadrupole 
moment $Q$, but a more intuitive quantity is a deformation parameter 
$\beta$, which removes the trivial dependence on the radius of a
nucleus from the quadrupole moment. In order to extract the 
deformation parameter from the quadrupole moment, one often 
uses the formula which is obtained by 
assuming a sharp-cut density distribution\cite{RS80,RGL97}. 
To partly account for the deviation of density distribution from a sharp-cut 
function, sometimes one also uses a formula in which the sharp-cut radius
is replaced by a root-mean-square radius (with a trivial constant
factor) \cite{SDNPD03,MDD00,YM00}. 
In general, the latter formula is believed 
to be better than the former when the surface diffuseness parameter 
of density distribution is large as in neutron-rich nuclei. 

In this paper, we systematically investigate 
whether that is the case. To this end, we first use 
a deformed Woods-Saxon density. This model has an advantage in that the
deformation parameter is given as an input parameter. This enables us
to check whether
the commonly used formulas lead to the correct value of deformation
parameter. We also derive the correction to the formulas due to a
finite value of surface diffuseness parameter. 
We then discuss the deformation parameter of neutron-rich 
Mg isotopes using the 
self-consistent relativistic mean-field (RMF) model. 

Let us begin with the definition for the intrinsic quadrupole moment, 
\begin{equation} 
Q=
\sqrt{\frac{16\pi}{5}}
\int d\vec{r}\,\rho(\vec{r})\,r^2Y_{20}(\theta). 
\label{Qmoment}
\end{equation} 
When the density $\rho(\vec{r})$ has a sharp edge, that is, 
\begin{equation}
\rho(\vec{r})=\rho_0\,\theta(R(\theta)-r),
\label{sharpdens}
\end{equation}
with $\rho_0=3A/(4\pi R_0^3)$, $A$ being the mass number of a
nucleus, and 
\begin{equation}
R(\theta)=R_0(1+\beta\,Y_{20}(\theta)),
\end{equation}
the quadrupole moment $Q$ is evaluated as, 
\begin{equation} 
Q =\sqrt{\frac{16\pi}{5}}\,\frac{3}{4\pi}\,AR_0^2\,\beta, 
\label{eq:constR}
\end{equation}
to the first order of deformation parameter $\beta$. 
One often takes $R_0=1.2\,A^{1/3}$ fm for the sharp-cut 
radius $R_0$ \cite{RGL97}. 
For a sharp-cut density (\ref{sharpdens}), the root-mean-square 
radius is calculated as 
\begin{equation}
\langle r^2\rangle = \frac{\int r^2\rho(\vec{r})\,d\vec{r}}
{\int \rho(\vec{r})\,d\vec{r}}\sim 
\frac{3}{5}\,R_0^2, 
\end{equation}
again to the leading order of $\beta$. 
Therefore, 
the relationship between the quadrupole moment and the deformation 
parameter given by Eq. (\ref{eq:constR}) can be also written as 
\cite{SDNPD03,MDD00,YM00} 
\begin{equation} 
 Q = \sqrt{\frac{16\pi}{5}}\frac{5}{4\pi}A \beta\,\langle r^2\rangle. 
\label{eq:rms}
\end{equation}

The effect of finite surface diffuseness of density distribution can
be accounted for using a deformed Woods-Saxon density, 
\begin{equation}
\rho(\vec{r})= \frac{\rho_0}{1+e^{(r-R_0-R_0\beta Y_{20}(\theta))/a}}. 
\label{eq:defWS}
\end{equation}
In order to derive the correction term to Eqs. (\ref{eq:constR}) and 
(\ref{eq:rms}), we expand Eq. (\ref{eq:defWS}) with respect to the
deformation parameter $\beta$ and keep only the first order term, that
is, 
\begin{equation}
\rho(\vec{r})
\sim \rho_0(r)-\frac{d\rho_0}{dr}\,R_0\beta Y_{20}(\theta)
\label{rhoexp}
\end{equation}
where 
\begin{equation}
\rho_0(r)=\frac{\rho_0}{1+e^{(r-R_0)/a}}. 
\label{WS}
\end{equation}
Notice that the angle dependence of the surface diffuseness parameter
$a$ \cite{BM75,HZ95,TTO96} 
as well as the dependence of $R_0$ on the deformation parameter
due to the volume conservation \cite{RS80} does not 
appear in Eq. (\ref{rhoexp}) since they are higher order terms of $\beta$. 
With the density given by Eq. (\ref{rhoexp}), the quadrupole moment
$Q$ is calculated as, 
\begin{equation}
Q = - \sqrt{\frac{16\pi}{5}}R_0\beta\int_{0}^{\infty} r^4 dr\,
\frac{d\rho_0}{dr}. 
\end{equation} 
Since $d\rho_0/dr$ has a finite value only in a small region 
near the nuclear surface $r\sim R_0$, we expand 
$r^4$ around $r = R_0$. Following the same procedure as in
Refs. \cite{BM69,R65}, we find 
\begin{equation}
Q \sim \sqrt{
\frac{16\pi}{5}}R_0^5\,\rho_0\,\beta
\left(1+2\pi^2\frac{a^2}{R_0^2}\right), 
\label{Q}
\end{equation} 
to the order of $(a/R_0)^2$. One can eliminate the dependence on 
$\rho_0$ in Eq. (\ref{Q}) using the condition for the
normalization, 
\begin{equation}
A=\int d\vec{r}\,\rho(\vec{r})\sim
\frac{4\pi}{3}R_0^3\,\rho_0\,
\left(1+\pi^2\frac{a^2}{R_0^2}\right). 
\end{equation}
This yields,
\begin{equation}
Q \sim 
\sqrt{\frac{16\pi}{5}}\frac{3}{4\pi}A\,R_0^2\beta
\,\left(1+\pi^2\frac{a^2}{R_0^2}\right).
\label{qmoment2}
\end{equation} 
Furthermore, one can use the relation 
\begin{equation}
\langle r^2\rangle \sim 
\frac{3}{5}\,R_0^2 
\left(1+\frac{7\pi^2}{3}\,\frac{a^2}{R_0^2}\right), 
\label{rms}
\end{equation}
to eliminate the dependence on $R_0$, leading to 
\begin{equation}
Q=
\sqrt{\frac{16\pi}{5}}\frac{3}{4\pi}A\beta
\,\left(\frac{5}{3}\langle r^2\rangle-\frac{4}{3}\pi^2
a^2\right).
\label{eq:diffu}
\end{equation}
A similar formula can be found also in Ref. \cite{BM75}. 

We now investigate the performance of 
Eqs. (\ref{eq:constR}), 
(\ref{eq:rms}), and (\ref{eq:diffu}) 
using realistic 
density distributions. To this end, we first compute the 
quadrupole moment (\ref{Qmoment}) using a deformed Woods-Saxon 
density, (\ref{rhoexp}). 
Following Ref. \cite{Chamon}, we choose 
$R_0 = 1.31A^{1/3}-0.84$ fm in Eq. (\ref{WS}). 
This value was obtained by fitting to theoretical as well as 
to experimental density distributions for a number of nuclei 
with the Woods-Saxon shape \cite{Chamon}. 
For a deformation parameter $\beta$ in Eq. (\ref{rhoexp}), we 
choose $\beta=0.3$. 
Once the quadrupole moment is obtained, we can use 
Eqs. (\ref{eq:constR}), (\ref{eq:rms}), and (\ref{eq:diffu}) to 
obtain an approximate value for the deformation parameter, $\beta$. 
If the formulas worked perfectly, they would lead to 
$\beta=0.3$ as in the original density distribution. 

Figure 1 shows the deformation parameters obtained in this way as a 
function of diffuseness parameter $a$ in the density distribution. 
The upper panel is for $A$=40, while the lower panel for $A$=238. 
The dashed line is obtained with Eq. (\ref{eq:rms}) using the
root-mean-square radius. We see that this formula 
significantly underestimates the deformation parameter, 
especially for the lighter system, $A$=40, 
except when 
the diffuseness parameter $a$ is close to zero. 
Surprisingly, the formula does not seem 
to work even for stable nuclei around $a\sim$ 0.55 fm. 
When one takes into account the surface diffuseness correction 
with Eq. (\ref{eq:diffu}), 
one obtains a reasonable agreement with the exact value 
of the deformation parameter, 
as shown by the solid line. 
The dot-dashed line is obtained by using Eq. (\ref{eq:constR}) with 
$R_0=1.2\,A^{1/3}$ fm. We see that this formula provides a reasonable 
value of the deformation parameter if the surface diffuseness
is around 0.55 fm as in stable nuclei, although the
deviation from the 
exact value becomes large when the surface diffuseness is 
around 1.0 fm. 

\begin{figure}[tb]
\begin{center}\leavevmode
\includegraphics[width=0.91\linewidth, clip]{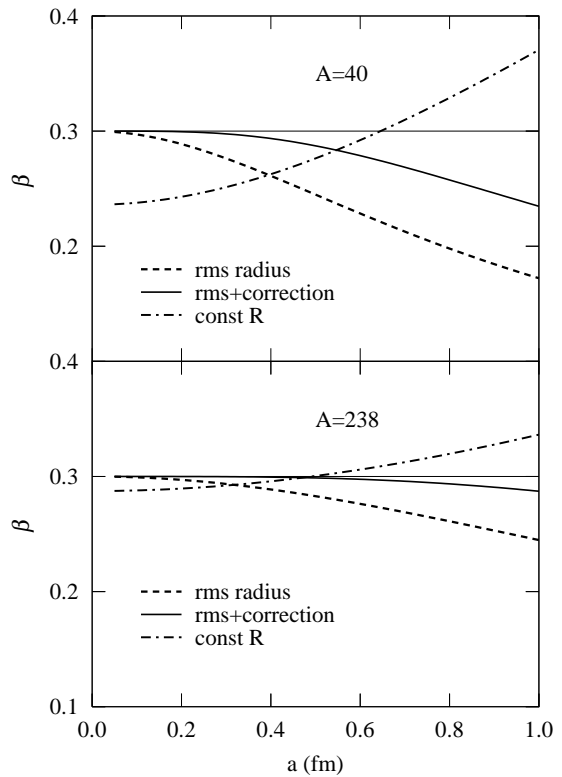}
\caption{
Deformation parameters obtained with several formulas as a function of 
surface diffuseness parameter $a$ in the density distribution. The
deformation parameter is set to be $\beta=0.3$ in the deformed
Woods-Saxon density. The upper panel is for $A$=40, while the lower 
panel for $A$=238. The dot-dashed line is obtained with
Eq. (\ref{eq:constR}) assuming a sharp-cut density, while the dashed
line with Eq. (\ref{eq:rms}) using the root-mean-square 
radius. The solid line is obtained with Eq. (\ref{eq:diffu}), 
including the surface diffuseness correction.} 
\label{fig:fermi}
\end{center}
\end{figure}

\begin{figure}[t]
\begin{center}
\includegraphics[width=0.91\linewidth, clip]{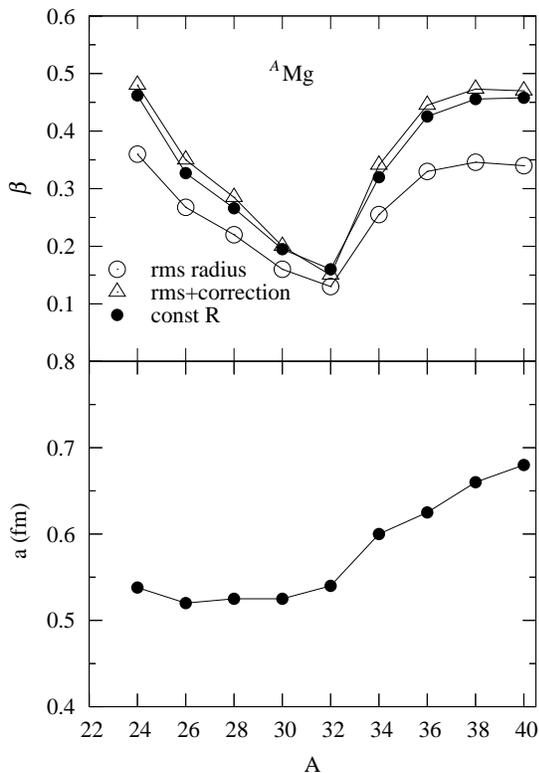}
\caption{Upper panel: Deformation parameters for Mg isotopes 
extracted from the 
relativistic mean-field (RMF) 
density distribution. 
The open triangles, the open circles, and 
the filled circles are obtained with Eqs. 
Eqs. (\ref{eq:constR}), (\ref{eq:rms}), and (\ref{eq:diffu}),
respectively. 
Lower panel: The diffuseness parameter for each nucleus obtained 
by fitting the RMF density to the Woods-Saxon shape. 
}
\label{fig:mg}
\end{center}
\end{figure}

Let us now discuss the deformation parameters obtained with
self-consistent RMF calculations. 
The upper panel of Fig. 2 shows the deformation parameters for 
Mg isotopes. 
The filled circles, the open circles, and 
the open triangles are obtained with  
Eqs. (\ref{eq:constR}), (\ref{eq:rms}), and (\ref{eq:diffu}),
respectively. 
In this calculation, we use the NL3 parameter set \cite{nl3} and 
assume the axial symmetry. 
We use the computer code RMFAXIAL \cite{RGL97}, 
which solves the RMF equations using the harmonic oscillator expansion 
method. We employ the constant gap approach with the pairing gap 
given in Refs. \cite{MN88,MN89}. In order to estimate the diffuseness 
parameter $a$, we expand the density distribution into multipoles 
(see Appendix B of Ref. \cite{Vautherin73}), 
\begin{equation}
\rho(r,\theta)=
\rho_0(r)+\rho_2(r)P_2(\cos\theta)+\rho_4(r)P_4(\cos\theta)+\cdots, 
\label{eq:multipole}
\end{equation}
and fit the monopole density $\rho_0(r)$ with the Woods-Saxon shape,
(\ref{WS}). The diffuseness parameter thus obtained is shown in the 
lower panel of Fig. 2. 
One finds that the 
deformation parameter estimated with 
Eq. (\ref{eq:rms}) (the open circles) 
is considerably smaller than that estimated with 
Eq. (\ref{eq:diffu}) which includes the surface diffuseness effect
(the open triangles), 
in accordance with the study with the deformed Woods-Saxon density. 
One also finds that the deformation parameters estimated with
Eq. (\ref{eq:constR}) are close to those with Eq. (\ref{eq:diffu})
(the filled circles), 
although it might simply be accidental. We have checked that our conclusions
remain the same also for Sr isotopes. 

\begin{figure}[t]
\begin{center}\leavevmode
\includegraphics[width=0.91\linewidth, clip]{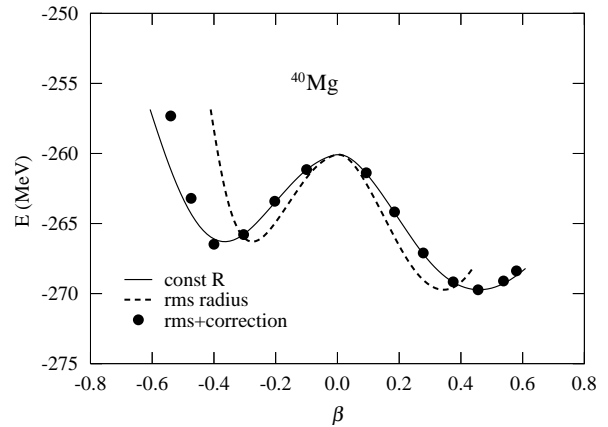}
\end{center}
\label{fig:surface}
\caption{Potential energy surface for $^{40}$Mg. 
The solid line, the dashed line, and the filled circles are 
obtained with  Eqs. (\ref{eq:constR}), 
(\ref{eq:rms}), and (\ref{eq:diffu}), respectively. }
\end{figure}

Figure 3 shows the potential energy surface 
for $^{40}$Mg nucleus as a function of deformation parameter $\beta$, 
obtained with constrained RMF calculations. 
The solid and the dashed lines are obtained with Eqs. (\ref{eq:constR})
and (\ref{eq:rms}), respectively. Since Eq. 
(\ref{eq:rms}) underestimates the deformation parameter, the
distance between the prolate and oblate minima is also
underestimated. The energy surface 
which takes into account 
the surface 
diffuseness correction is denoted by the filled circles. As we showed
in Fig. 2, it is similar to the energy surface obtained with
Eq. (\ref{eq:constR}). 

In attempting to extract the deformation parameter from an experimental
quadrupole moment, 
an interesting question is how to estimate the surface diffuseness 
parameter $a$ from experimental data. 
Unfortunately, the root-mean-square radius alone does
not determine the radius and the surface diffuseness parameters, $R_0$
and $a$, simultaneously. 
One possible prescription is to assume 
$R_0 = 1.31A^{1/3}-0.84$ fm as in Ref.\cite{Chamon}, and use
Eq. (\ref{rms}) to estimate the surface diffuseness parameter $a$ 
from the root-mean-square radius. 
In fact, our RMF calculations 
show that the $R_0$ parameter estimated from the fitting to the
monopole density is well parametrized by this function up to the drip line 
nucleus, 
at least for the Mg isotopes presented in Fig. 2. 
One can then use Eq. (\ref{qmoment2}) to estimate
the deformation parameter $\beta$ from the quadrupole moment $Q$. 
This procedure may be important in discussing the deformation
parameter of neutron-rich nuclei, where 
the surface diffuseness 
parameter is expected to be significantly larger than that of stable
nuclei\cite{DHNS94}. 

In summary, 
we discussed 
the role of surface diffuseness parameter of density
distribution in converting the quadrupole moment to 
the deformation parameter. 
For this purpose, we used both the deformed Woods-Saxon and the RMF 
density distributions. 
We showed that the widely used linear order formula 
with root-mean-square radius significantly underestimates 
the deformation parameter. 
After including the surface diffuseness correction, the 
resultant deformation parameters were found to be close to 
those estimated with the 
linear order formula with a sharp-cut radius. 

The present consideration will be important when one discusses the 
deformation properties of neutron-rich nuclei, where 
the surface diffuseness 
parameter is expected to be large. 
In particular, 
when one draws a two dimensional potential energy surface 
spanned by 
proton and neutron 
deformation parameters, $\beta_p$ and $\beta_n$, it may look 
considerably different depending on which
formula one employs in estimating the deformation parameters. 
Such studies are now in progress, and 
we will report on them in a separate paper \cite{LHT06}. 
Also, it will be an interesting future problem to extend the formula
derived in this paper by including the higher order terms of
deformation parameter $\beta$. This will involve the angle dependent
surface diffuseness parameter and the condition for volume
conservation. 

\bigskip
We thank H. Sagawa for 
useful discussions. 
This work was supported by the Grant-in-Aid for Scientific Research,
Contract No. 16740139 from the Japanese Ministry of Education,
Culture, Sports, Science, and Technology.


\begin{thebibliography}{99}

\bibitem{RS80}
P. Ring and P. Schuck, {\it The Nuclear Many Body Problem}
(Springer-Verlag, New York, 1980).

\bibitem{RGL97}P. Ring, Y.K. Gambhir, and G.A. Lalazissis, 
Comp. Phys. Comm. {\bf 105}, 77 (1997); 
Y.K. Gambhir, P. Ring, and A. Thimet, Ann. of Phys. (N.Y.), 
{\bf 198}, 132 (1990). 

\bibitem{SDNPD03}M.V. Stoitsov, J. Dobaczewski, W. Nazarewicz,
  S. Pittel, and D.J. Dean, Phys. Rev. C{\bf 68}, 054312 (2003). 

\bibitem{MDD00}H. Molique, J. Dobaczewski, and J. Dudek, 
Phys. Rev. C{\bf 61}, 044304 (2000). 

\bibitem{YM00}M. Yamagami and K. Matsuyanagi, 
Nucl. Phys. {\bf A672}, 123 (2000). 

\bibitem{BM75}A. Bohr and B. Mottelson, {\it Nuclear Structure} (Benjamin,
New York, 1975), vol. 2, Eqs. (4-188c) and (4-191). 

\bibitem{HZ95}I. Hamamoto and X.Z. Zhang, Phys. Rev. C{\bf 52}, R2326
(1995). 

\bibitem{TTO96}N. Tajima, S. Takahara, and N. Onishi, 
Nucl. Phys. {\bf A603}, 23 (1996). 
 
\bibitem{BM69}A. Bohr and B. Mottelson, {\it Nuclear Structure} (Benjamin,
New York, 1969), vol. 1, p. 160. 

\bibitem{R65}F. Reif, {\it Fundamentals of Statistical and Thermal
  Physics} (McGraw Hill, New York, 1965), p. 394. 

\bibitem{nl3}G.A. Lalazissis, D. Vretenar, and P.Ring, Phys. Rev
  C{\bf 55}, 540 (1997)

\bibitem{MN88}P. M\"oller and J.R. Nix, At. Data Nucl. Data Table {\bf
  39}, 225 (1988). 

\bibitem{MN89}D.G. Madland and J.R. Nix, Nucl. Phys. {\bf A476}, 1
  (1989). 

\bibitem{Vautherin73}D. Vautherin, Phys. Rev. C{\bf 7}, 296, (1973)

\bibitem{Chamon}L.C. Chamon, B.V. Carlson, L.R. Gasques, D. Pereira,
C. De Conti, M.A.G. Alvarez, M.S. Hussein, M.A. Candido Ribeiro,
E.S. Rossi, Jr., and C.P. Silva, Phys. Rev C{\bf 66}, 014610, (2002). 

\bibitem{DHNS94}J. Dobaczewski, I. Hamamoto, W. Nazarewicz, and J.A. 
Sheikh, Phys. Rev. Lett. {\bf 72}, 981 (1994). 

\bibitem{LHT06}N.W. Lwin, K. Hagino, and N. Takigawa, to be 
published. 




\end{thebibliography}
\end{document}